\documentclass[11pt]{article}
\usepackage{epsfig}
\usepackage{amsmath}

\newlength{\dinwidth}
\newlength{\dinmargin}
\title{\bf  The QCD odderon in elastic (anti)proton scattering}

\author{M.A.Braun\\
Dep. of High Energy physics,
Saint-Petersburg State University,\\
198504 S.Petersburg, Russia}

\newcommand{\beq}{\begin{equation}}
\newcommand{\eeq}{\end{equation}}
\newcommand{\mbf}[1]{\mbox{\boldmath $#1$}}
\newcommand{\ba}{\bar{k}_1}
\newcommand{\bb}{\bar{k}_2}
\newcommand{\bh}{\bar{h}}

\newcommand{\kk}{\mbf{k}}

\begin{document}
\maketitle
\begin{abstract}
\noindent
The C-odd amplitude for the elastic $pp$ and $p\bar{p}$ scattering
due to the exchange of the QCD odderon proposed by J.Bartels, L.N.Lipatov and G.P. Vacca
is calculated with the Fukugita-Kwiecinski proton impact factor. The found amplitude is very small and
cannot be felt in the differential cross-sections at 2.76 and 1.96 Tev respectively.
\end{abstract}
\section{Introduction. The perturbative QCD odderon}
The perturbative QCD odderon exists in two different states. The first one
was proposed as the $C=-1$ eigenstate of the Hamiltonian
for three reggeized gluons (the simplest of the series of the so-called BKP
states ~\cite{BKP1,BKP2}). Its properties and conformal invariance were discussed 
very long ago ~\cite{lipatov}. After several attempts to numerically
estimate its intercept ("energy") and wave function these was finally found by Janik and
Wosiek ~\cite{jawos}. The found intercept turned out to be below unity
\[ \alpha_0^{jw}=1-0.24717\bar{\alpha},\ \ \bar{\alpha}=\frac{\alpha_s N_c}{\pi}.\]
Later another odderon state was constructed by J.Bartels, L.N.Lipatov and G.P.Vacca as a  degenerate
3-reggeon state with 2 reggeons located at the same spatial point. Ir is described by
solutions of the BFKL equation for the pomeron with odd conformal spins ~\cite{BLV}. Accordingly
the vaximal intercept of its groundstate is exactly unity. So inevitably at very high energies this BLV odderon dominates.

Possible manifestations of the odderon in the experiment include first of all a series of processes
which can occur exclusively by the $C=-1$ exchanges, such as transitions $\gamma$ (or $\gamma*)$ to $\eta_c$.
Numerical estimates of the corresponding probabilities have given very small values, which practically prohibit these
direct searches for the odderon. Another possibility, much discussed recently in view of the current experimental data,
is to look for the odderon in the difference between $pp$ and $p\bar{p}$ elastic cross-sections. In these  reactions the
odderon exchange enters with a different sign and in the cross-sections it is multiplied by the leading and big $C=+1$
exchange. So the above difference contains the odderon exchange linearly and not quadratical in contrast to reactions
realizable only by the $C=-1$ exchange. This raises some hopes to see the odderon more easily. The recent experimental data
seem to exhibit a definite difference between $pp$ and $p\bar{p}$ cross-sections and so
a presence of the odderon exchange ~\cite{pp276,yes2,yes3,yes4}, although there are certain doubts on this point~\cite{doubt1,doubt2}.

Actually the role of the perturbative odderon in the $pp$ and $p\bar{p}$ scattering was studied long ago in the approach
in which gluon interactions inside the odderon were neglected and the odderon was considered as just the three gluon exchange
~\cite {DL}. Later  the problem of the nonperturbative proton impact factor was discussed ~\cite{ewerz}. The conclusion
of these earlier papers was quite optimistic: with a suitable choice of the QCD coupling constant use of the three gluon exchange for the $C=-1$
amplitude lead to quite good agreement with the experimental data existing at that moment: $\sqrt{s}<62.5$ GeV for $pp$
scattering and $\sqrt{s}=53$ GeV for $p\bar{p}$ scattering. The authors of ~\cite{ewerz} pointed out the importance of taking account
the gluon interactions inside the reggeon.

In this paper we present a partial solution of this problem considering the BLV oddderon exchange instead of the simple triple gluon one.
 We calculate the $C=-1$ $pp$ and $p\bar{p}$ amplitudes due to the interaction with  the BLV odderon
and putting it together with the $C=+1$ amplitude find the final cross-section for the two elastic processes.
Of course the immediate question is from where we can take the $C=+1$ amplitude. In absence of any trustful theory, as long ago,
we can use only phenomenological amplitudes. In this study we use two  models which claim to
successfully describe the data
up to $\sqrt{s}=7$ TeV.  The first, proposed in ~\cite{broni}, is especially convenient for our purpose, since it is
based on the Regge description of different contributions to the amplitude. The second ~\cite{silva}, although indirectly also
based on the Regge approach, does not distinguish between Regge components. However with different parametrizations it
describes both the $pp$ and $p\bar{p}$ amplitudes and so allows to extract the desired $C=+1$ amplitude.

In the theoretical $C=-1$ amplitude the BLV odderon is attached to two (anti)proton impact factors, which are unperturbative and
so model-dependent We use the impact factor proposed by Fukugita and Kwiecinski based on the perturbative picture for
the interaction of three quarks with three gluons ~\cite{FK}:
\beq
\Phi_p=d\Big[F(\mbf{q},0,0)-\sum_{i=1}^3F(\mbf{k}_i,
\mbf{q}-\mbf{k}_i,0)+2F(\mbf{k}_1,\mbf{k}_2,\mbf{k}_3)\Big],
\label{if}
\eeq
where
\beq
F(\mbf{k}_1,\mbf{k}_2,\mbf{k}_3)=
\frac{2a^2}{2a^2+(\mbf{k}_1-\mbf{k}_2)^2+
(\mbf{k}_2-\mbf{k}_3)^2+(\mbf{k}_3-\mbf{k}_1)^2},
\label{Fp}
\eeq
with
$d=8(2\pi)^2g_p^3$
and the scale parameter $a= m_{\rho}/2$.
The impact factor (\ref{if}) satisfies the basic requirement
that it should vanish when any of the three gluon momenta goes to zero.
It is proportional to $g_p^3$
where $g_p$ is an effective and so unknown QCD coupling constant inside the proton. So strictly speaking the magnitude
of the odderon-(anti)proton coupling is unknown and is in fact an arbitrary parameter.
From the comparison with the two gluon exchange model for hadronic
cross-sections the authors of ~\cite{Kwie} estimated
$\alpha_p=g_p^2/4\pi\simeq 1$.

Our calculations show that the gluon interactions responsible for the formation of the BLV odderon strongly diminish the
oddderon amplitude (around 1000 times). With $\alpha_p=1$ the odderon exchange turns out to be far below any significant effect in the
$pp$ or $p\bar{p}$ scattering. To obtain results which more or less agree with the experimental contribution of the $C=-1$
component of the relevant amplitudes one has to augment the value of $\alpha_p $ from unity to $\sim$ 14, which does not seem
reasonable.

As we noted our result only partially resolves the QCD odderon problem in the $pp$ and $p\bar{p}$ elastic scattering.
The remaining task is to study the different JW odderon, which is  made of three reggeons at different spatial
points. This is a much more difficult question since the total spectrum of the JW odderon states and so its Green function
remain unknown and the relevant technical problems seem great. This is a problem for  future studies.

\section{$pp$ and $p\bar{p}$ elastic scattering. Phenomenological description}
We use the normalization in which
 the differential cross-section for $pp$ and $p\bar{p}$ scattering is given by the formula
\beq
\frac{d\sigma}{dt}(p(\bar{p})+p\rightarrow p(\bar{p})+p)=
\frac{\pi}{ s^2}|{\cal A}|^2.
\label{cross1}
\eeq
Here ${\cal A}(s,t)$, is corresponding amplitude.
which splits into the sum or difference of its $C=+1$ and $C=-1$ parts
\beq
{\cal A}_{pp}^{p\bar{p}}={\cal A}_+\pm{\cal A}_-.
\label{apm}
\eeq.
The odderon contribution  included into ${\cal A}_-$ is given by a convolution
of the two  proton impact factors  $\Phi_p$ and
the Odderon Green function $G_3$
\beq
{\cal A}_{odd}=\frac{s}{128\pi}\frac{5}{6}\frac{1}{3!}\frac{1}{(2\pi)^8}
\langle \Phi_p|G_3|\Phi_p\rangle.
\label{ampli}
\eeq
Here the matrix element is
\beq
\langle \Phi_p|G_3|\Phi_p\rangle=\int d\mu(\kk)\int d\mu(\kk')\Phi_p^*(\{\kk_i\})G_3(y,\{\kk_i\}.\{\kk_i'\})\Phi_p(\{\kk'_i\})
\label{mel}
\eeq
and the measure is $d\mu(\kk)=d^2k_1d^2k_2d^2k_3\delta^3(k_1+k_2+k_3-q)$ where $q^2=-t$.

We assume that the c.m. energy squared $s$ is very high, so that
presumably in the in odd amplitude all contributions except for the odderon, coming from other
exchanges have practically died out.  This fact is confirmed by phenomenological descriptions
proposed for energies above 546 GeV.

 In the literature we have found two such descriptions, which on the one hand successfully describe the data up to 7 TeV and on the other
hand allow for the separation of the $C-1$ amplitude.

The first one was proposed in~\cite{broni} in 2018. It presented  ${\cal A}_\pm$ as a sum of contributions from different Regge exchanges.
The pomeron and odderon were taken as dipole Regge singularities. The pomeron contribution was taken as
\beq
{\cal A}_P=i\frac{a_Ps}{b_Ps_{0P}}\Big(r^2_{1P}(s)e^{r^2_{1P}(s)(\alpha_P-1)}- \epsilon_P r^2_{2P}(s)e^{r^2_{2P}(s)(\alpha_P-1)}\Big),
\label{broni}
\eeq
where
\[ r_{1P}(s)=b_P+l_P-i\frac{\pi}{2},\ \ r_{2P}=l_P-i\frac{\pi}{2},\ \ l_P=\ln\frac{s}{s_{0P}}\]
and the pomeron trajectory $\alpha_P$
\[
\alpha_P(t)=1+\Delta_P+\alpha'_Pt.
\]
It contained 6 parameters: $a_P\,,b_P\,,\Delta_P\,,\alpha'_P\,,\epsilon_P$ and $s_{0P}$.
The odderon contribution was taken in the same form with new parameters
$a_O\,,b_O\,,\Delta_O\,,\alpha'_O\,,\epsilon_O$ and $s_{0O}$.
Apart from the pomeron the ${\cal A}_+$ amplitude was taken to have a contribution from the $f$ meson
\[
{\cal A}_f= a_fe^{-i\pi \alpha_f(t)/2+b_ft} \Big(\frac{s}{s_{0f}}\Big)^{\alpha_f(t)}
\]
with $\alpha_{f}(t)=\alpha_{f0}+\alpha'_{f0}t$.
It contained 5 parameters $a_f\,,b_f\,,\alpha_{0f}\,,\alpha'_{0f}$ and $s_{0f}$.
The odd amplitude apart from the odderon was assumed to have a contribution from the $\omega$ meson
of the same form with parameters $a_\omega\,,b_\omega\,,\alpha_{0\omega}\,,\alpha'_{0\omega}$ and $s_{0\omega}$.
From the total set of 26 parameters 7 were fixed on physical grounds and the rest were fitted to the existing experimental data
on the differential elastic $pp$ and $p\bar{p}$ cross-sections as well as to the data on the total cross-section and parameter $\rho$.
One can find the values of the fitted parameters in ref. ~\cite{broni}.

The second description was proposed in ~\cite{silva} in 2019. It was based on the modified Phillips-Barger model ~\cite{fagundes}
in which the scattering amplitude was parametrized as follows
\beq
{\cal A}(s,t)=i\Big[F_p^2\sqrt{A}e^{Bt/2}+e^{i\phi}\sqrt{C}e^{Dt/2}\Big].
\label{pb}
\eeq
Here $F_p$ is the Dirac form-factor of the proton.
It contains a set of only 5 parameters, different for $pp$ and $p\bar{p}$ scattering.
They all depend on the energy. In ~\cite{silva} an interpolation of the $pp$ parameters
was proposed for energies in the range from 25 Gev to 13 TeV as quadratic functions
of $\ln s$. For the $p\bar{p}$ scattering two sets of parameters were given for energies
546 GeV and 1.8-1.96 TeV.
This parametrization does not allow to directly separate  the odderon exchange amplitude but
rather the total $C=-1$ amplitude as ${\cal A}_O={\cal A}_{p\bar{p}}-{\cal A}_{pp}$.
However at energies of the order 2 TeV one expects that the contribution from all $C=1$
exchanges other than the odderon are insignificant, so that one can identify ${\cal A}_-$ with the odderon exchange.

\section{The BLV  odderon}
The  QCD  BLV Odderon
was found in ~\cite{BLV}. Its properties and coupling to the proton impact factor
were discussed in some details in ~\cite{BBCV}. Here we reproduce some main points necessary for
understanding our calculations.

The odderon wave function in the 3-gluon momentum space is constructed
from the
known pomeron solutions $E^{(\nu,n)}$ ~\cite{lip1},
with odd $n=\pm 1,\pm 3,...$. Their intercept $\chi$ quickly goes down with $|n|$ so we shall be interested
only in $n=\pm 1$
\beq
\chi(\nu,\pm 1)=-2\zeta(3)\bar{\alpha}\nu^2,\ \ \bar{\alpha_s}=\frac{N_c\alpha_s}{\pi}.
\label{eigen}
\eeq
It was demonstrated in ~\cite{BLV} that
\beq
\Psi^{(\nu,\pm 1)}(\mbf{k}_1,\mbf{k}_2,\mbf{k}_3)=c(\nu)
\sum_{(123)} \frac{(\mbf{k}_1+\mbf{k}_2)^2}{\mbf{k}_1^2\mbf{k}_2^2}
E^{(\nu,\pm 1)}(\mbf{k}_1+\mbf{k}_2,\mbf{k}_3),\ \ c=\frac{1}{4\nu\sqrt{10\pi\zeta(3)}}
\label{oddwave}
\eeq
satisfies the odderon equation and has  intercept (\ref{eigen}).

Function $E^{(\nu,\pm 1)}(\mbf{k}_1,\mbf{k}_2)$ is the Fourier transform of the well-known
BFKL eigenfunctions
\beq
E^{(h,\bar{h})}(\mbf{r}_{10},\mbf{r}_{20})=
\left(\frac{r_{12}}{r_{10}r_{20}}\right)^h
\left(\frac{\bar{r}_{12}}{\bar{r}_{10}\bar{r}_{20}}\right)^{\bar{h}},
\label{pom_coord}
\eeq
where $\mbf{r}_{10}=\mbf{r}_1-\mbf{r}_0$ etc,
$h=(1+n)/2+i\nu$,
$\bar{h}=(1-n)/2+i\nu$, $n=\pm 1$ and the standard complex notation for
two-dimensional vectors is used on the right-hand side.

The Green function $G_3$ corresponding to the propagation of the BLV odderon
turns out to be given by
\[
G_3(y|\mbf{k}_1,\mbf{k}_2,\mbf{k}_3|\mbf{k}'_1,\mbf{k}'_2,\mbf{k}'_3)=\]
\beq
\sum_{n=\pm 1}\int_{-\infty}^{+\infty} d\nu e^{y\, \chi(\nu,n)}
\frac{(2\pi)^2(\nu^2+1/4)}{\nu^2(\nu^2+1)}
\Psi^{(\nu,n)}(\mbf{k}_1,\mbf{k}_2,\mbf{k}_3)
{\Psi^{(\nu,n)}}^*(\mbf{k}'_1,\mbf{k}'_2,\mbf{k}'_3),
\label{greenf}
\eeq
where $\Psi^{(\nu,n)}(\mbf{k}_1,\mbf{k}_2,\mbf{k}_3)$ are given by (\ref{oddwave})
and $y=\ln(s/s_0)$ is the rapidity. In (\ref{greenf}) one sees $\nu^2$ in the denominator.
However, as we shall presently find  this $\nu^2$ will  be fully canceled by the $\nu^4$ coming from
the product of the proton- odderon couplings. As a result the integrand of (\ref{oddwave}
behaves as $\nu^2$ at $\nu\to 0$.

The matrix element in (\ref{ampli}) becomes
\beq
\langle \Phi_p|G_3|\Phi_p\rangle=
\sum_{n=\pm 1}\int_{-\infty}^{+\infty} d\nu e^{y\, \chi(\nu,n)}
\frac{(2\pi)^2(\nu^2+1/4)}{\nu^2(\nu^2+1)}
\Big|\Big<\Phi_p|\Psi^{(\nu,n)}\Big>\Big|^2,
\label{mel1}
\eeq
where
\beq
\Big<\Phi_p|\Psi^{(\nu,n)}\Big>=\int d\mu(\kk)\Phi_p(\{\kk_i\})\Psi{(\nu,n)}(\{\kk_i\})
\label{mel2},
\eeq
or using (\ref{oddwave})
\beq
\Big<\Phi_p|\Psi^{(\nu,\pm 1)}\Big>=\frac{3}{4\nu\sqrt{10\pi\zeta(3)}}J(q).
\label{mel4}
\eeq
Here
\beq
J(q)=\int d^2lf({\bf l})E^{(\nu,\pm 1)}({\bf l},{\bf q}-{\bf l}),
\label{mel3}
\eeq
where
\beq
f({\bf l})=\int d^2k\frac{l^2}{k^2({\bf l}-{\bf k})^2} \Phi_p(\kk_1,\kk_2,\kk_3)
\label{fl}
\eeq
with
$\mbf{k}_1=\mbf{k}$, $\mbf{k}_2=\mbf{l}-\mbf{k}$ and $\mbf{k}_3=
\mbf{q}-\mbf{l}$.

Performing Fourier transformation one obtains the explicit expression for
$E^{(h\bh)}(\mbf{k}_1,\mbf{k}_2)$ ~\cite{BBCV}
\beq
E^{(h\bh)}(\mbf{k}_1,\mbf{k}_2)=C\Big(X(\mbf{k}_1,\mbf{k}_2)+
-X(\mbf{k}_2,\mbf{k}_1)\Big),\ \ C=\frac{\nu}{(4\pi)^2}(1+i\nu)\Gamma(1-i\nu)\Gamma(2--i\nu)
\label{ek}
\eeq
where
$X$ is expressed via hypergeometric functions. For $n=1$
\beq
X(\mbf{k}_1,\mbf{k}_2)=\left(\frac{k_1}{2}\right)^{i\nu-2}
\left(\frac{\bb}{2}\right)^{i\nu-1}
F\Big(-i\nu,1-i\nu;2;-\frac{\ba}{\bb}\Big)
F\Big(1-i\nu,2-i\nu;2;-\frac{k_2}{k_1}\Big)
\label{xx}
\eeq

The behavior of $\chi(\nu)$ indicates that at large $y$ the contribution comes from the region
of small $\nu$. This means that in the limit of large $y$ it is sufficient to know function
$E^{(\nu,\pm 1)}(\mbf{k}_1,\mbf{k}_2)$ at small values of $\nu$. One obtains in the first order
in $\nu$
\beq
E_1(\mbf{k}_1,\mbf{k}_2)=\frac{\nu}{2\pi^2 q}\left(\frac{1}{k_1\bb}-
\frac{1}{k_2\ba}\right).
\eeq
This function is antisymmetric in the azimuthal angle. So it is orthogonal
to the two impact factors which are azimuthal symmetric.
For this reason a non-zero contribution only comes from the terms
quadratic in $\nu$. Omitting those of them which have the same structure
as (23) we find
\beq E_2(\mbf{k}_1,\mbf{k}_2)=
\frac{i\nu^2}{2\pi^2q}\Big[
\frac{1}{\mbf{k}_2^2}\ln \mbf{k}_1^2-\frac{1}{\mbf{k}_1^2}\ln \mbf{k}_2^2
+\left(\frac{1}{\mbf{k}_1^2}-\frac{1}{\mbf{k_2}^2}\right)\ln q^2\Big].
\label{e2}
\eeq
As a result the matrix element (\ref{mel3}) behaves as $\nu^2$ at small $\nu$. Its square
 gives $\nu^4$, which converts $\nu^2$ in the denominator of (\ref{mel1}) to $\nu^2$ in the numerator.

Leaving in the integrand of (\ref{mel1}) only the exponential factor multiplied by $\nu^2$ and performing
integration over $\nu$ and summation over $n=\pm1$ one finally obtains
\beq
\Big<\Phi_p|G_3|\Phi_p\Big>=by^{-3/2}J^2(q),\ \
b=\frac{9}{320\sqrt{2\pi}}{\bar{\alpha}}^{-3/2}\zeta(3)^{-5/2}
\label{mel5}
\eeq
It diminishes with energy as $y^{-3/2}$.

\section{The impact factor and the final odderon amplitude}
The proton impact factor is non-perturbative. We use the Fukugita-Kwiecinski impact factor
(\ref{if},\ref{Fp})
proposed in ~\cite{FK} and used in ~\cite{Kwie, BBCV}
with $\alpha_p=1$.
As mentioned the impact factor (\ref{if}) vanishes
when any of the three gluon momenta goes to zero.
This guarantees that calculation of $f({\bf l})$ given by (\ref{fl}) is
infrared convergent.

Explicitly one finds
\beq
f(\mbf{l})= \int d^2\mbf{k}\, \frac{\mbf{l}^2}{\mbf{k}^2(\mbf{l}-\mbf{k})^2}
\big[F(\mbf{q},0,0)-\sum_{j=1}^3F(\mbf{k}_j,\mbf{q}-\mbf{k}_j,0)
+2F(\mbf{k}_1,\mbf{k}_2,\mbf{k}_3)\Big]
\label{funprot}
\eeq
where $\mbf{k}_1=\mbf{k}$, $\mbf{k}_2=\mbf{l}-\mbf{k}$ and $\mbf{k}_3=
\mbf{q}-\mbf{l}$.
The  integral (\ref{funprot}) is infrared finite, since the
square bracket vanishes if any of the gluon momenta go to zero. However,
individual terms inside the square bracket are infrared divergent. So
at intermediate stages it is convenient to introduce an auxiliary infrared regularization.
The integral $J(q)$ given by (\ref{mel3}) contains 4 integrations. One of them in
$f({\bf l})$ can be done analytically due to the simple form of $\Phi_p$
(see ~\cite{BBCV} for details). The other three require numerical integration.

The final odderon amplitude ${\cal A}_O$ depends on the two coupling constants $\bar{\alpha}$
and $\alpha_p$, the latter referring to the unperturbative coupling inside the proton.
In fact only the overall magnitude of the amplitude depends on them: it is
proportional to $\alpha_p^3{\bar{\alpha}}^{-3/2}$. In our calculations we fixed $\bar{\alpha}=0.2$.

\begin{figure}
\begin{center}
\epsfig{file=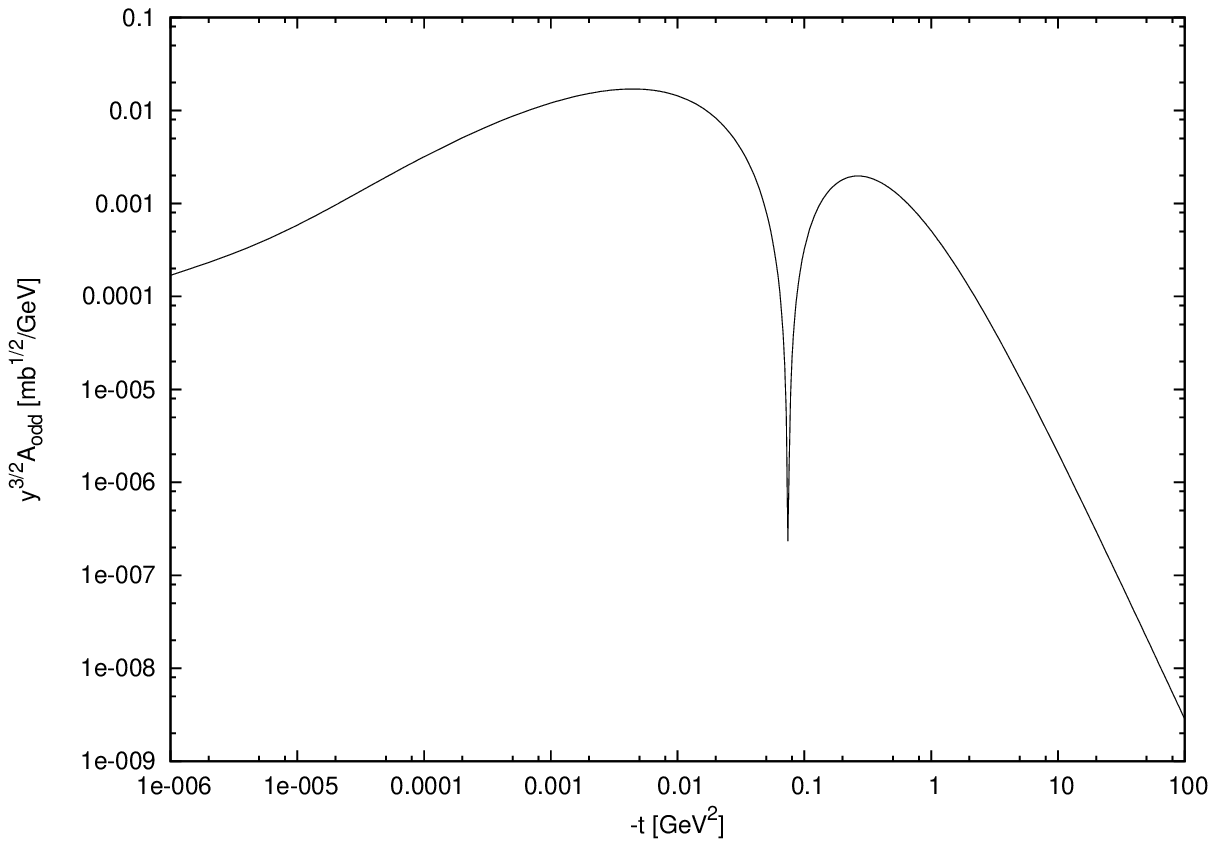, width=0.45\columnwidth }
\epsfig{file=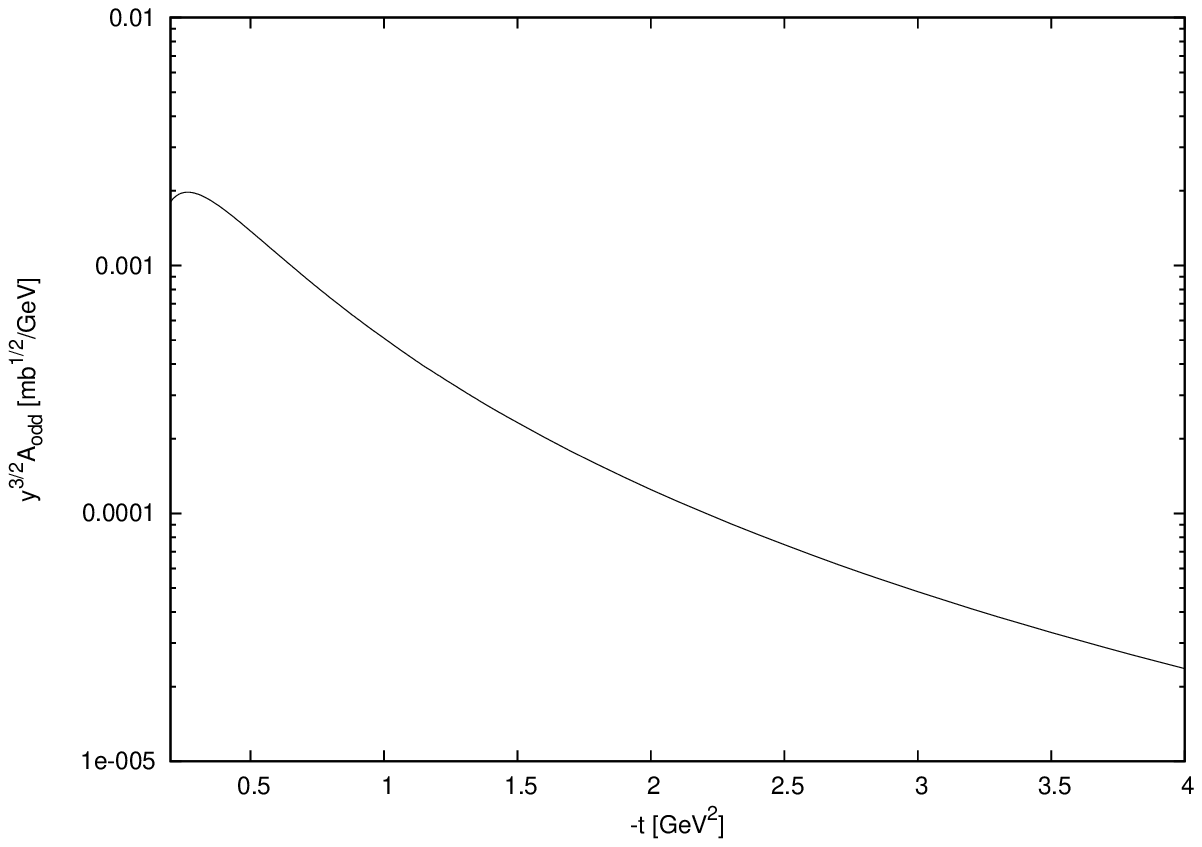, width=0.45\columnwidth }
\end{center}
\caption{The calculated odderon amplitude ${\cal A}_O$ with the proton coupling $\alpha_p=1$
 multiplied by $y^{3/2}$ for $0<-t<100$ GeV$^2$ in the logarithmic scale  (left panel) and for
 $0.2<-t<4$ GeV$^2$ in the natural scale (right panel)}
\label{fig1}
\end{figure}

With the original value $\alpha_p=1$ our results are presented in Fig.\ref{fig1}.
To avoid energy dependence of the plot we actually show $y^{3/2}{\cal A}_O$ which is
energy independent. The two panels in Fig. \ref{fig1} illustrate on the one hand the $t$ dependence in the
whole region $0<-t<100$ GeV$^2$ with particular attention to the behavior at very small $|t|$ and on the other
the $t$ dependence in the region $0.2<-t< 4$ GeV$^2$ relevant for the experimental setup.
As we observe the odderon amplitude exhibits a rather whimsical behavior in $t$. At $t=0$ it goes to zero
as $\propto |t|$. So it does not contribute to the ratio $\rho={\rm Re}{\cal A}/{\rm Im}{\cal A}$ at $t=0$,
which is important in relation to experimental bservations (see ~\cite{yes2,doubt1}).

\begin{figure}
\begin{center}
\epsfig{file=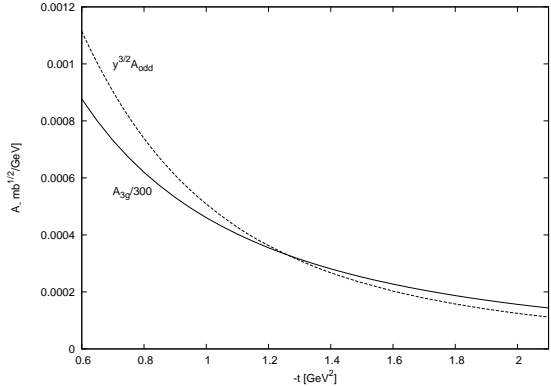, width=0.45\columnwidth }
\end{center}
\caption{The calculated odderon amplitude ${\cal A}_O$ with the proton coupling $\alpha_p=1$
 multiplied by $\sqrt{y}$ for $0.6<-t<2$ GeV$^2$  compared with the exchange of three non-interacting gluons in the $C=-1$
 state with the same $\alpha_p$}
\label{fig2}
\end{figure}

To study the influence of the gluon interaction in the 3-gluon exchange we also calculated the $C=-1$
amplitude corresponding to the non-interacting three-gluon exchange, used in the old paper \cite{ewerz}
for low energies with very optimistic conclusions. The three gluon exchange is given by the same formula
(\ref{ampli}) in which the matrix element is just
\beq
\langle \Phi_p|G_3|\Phi_p\rangle_{3g}=\int d\mu(\kk)\Big|\Phi_p(\{\kk_i\})\Big|^2\frac{1}{k_1^2k_2^2k_3^2}.
\label{mel6}
\eeq
In this case all 4 integrations have to be performed numerically.
Our results show that, first, this amplitude ${\cal A}_{3g}$
shows a smooth behavior in $t$. At $t=0$ it is finite and equal to  4.29 $\sqrt{mb}/GeV$. With the growth if $|t|$
it monotonously diminishes. Second,
${\cal A}_{3g}$  is much greater than the odderon amplitude ${\cal A}_O$.
With the same $\alpha_p$ three-gluon exchange amplitude is roughly 300 times greater than
$y^{3/2}{\cal A}_O$. This is illustrated in Fig.\ref{fig2} where we compare ${\cal A}_{3g}/300$
and $y^{3/2}{\cal A}_O$. Note that in ~\cite{ewerz} the pomeron coupling constant $\alpha_p$ was
equal to 0.3, which means that their ${\cal A}_{3g}$ was 27 times smaller than in Fig. \ref{fig2}
However this still remains far above the odderon amplitude with $\alpha_p=1$.
So  it turns out that gluon interactions drastically diminish the three-gluon exchange in the BLV odderon.

\section{Cross-sections}
In this section we study the cross-sections which are obtained with the found odderon amplitude an compare them
with the excising experimental data.
The $pp$ data cover a wide energetic interval from low energies up to 13 TeV. Unfortunately the $p\bar{p}$ data are much more scarce.
At high energies we shall consider both data at the closest possible  energies: $pp$ at 2.76 GeV and $p\bar{p}$ at 1.96 GeV.
The two models discussed in the Section 1 give possibilities to present the relevant odderon amplitudes (in ~\cite{broni}
and in ~\cite{silva} assuming  in the latter case that all other reggeon exchanges are insignificant).

\begin{figure}
\begin{center}
\epsfig{file=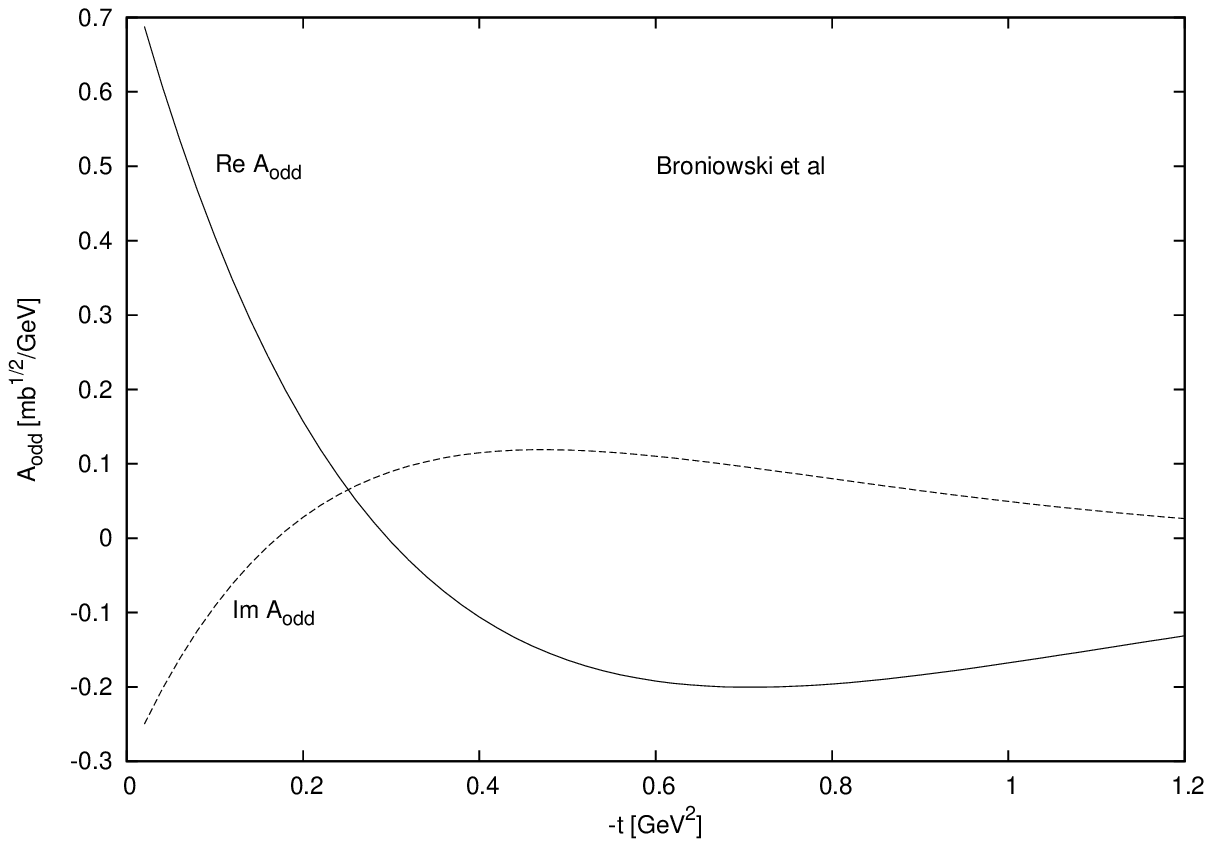, width=0.45\columnwidth }
\epsfig{file=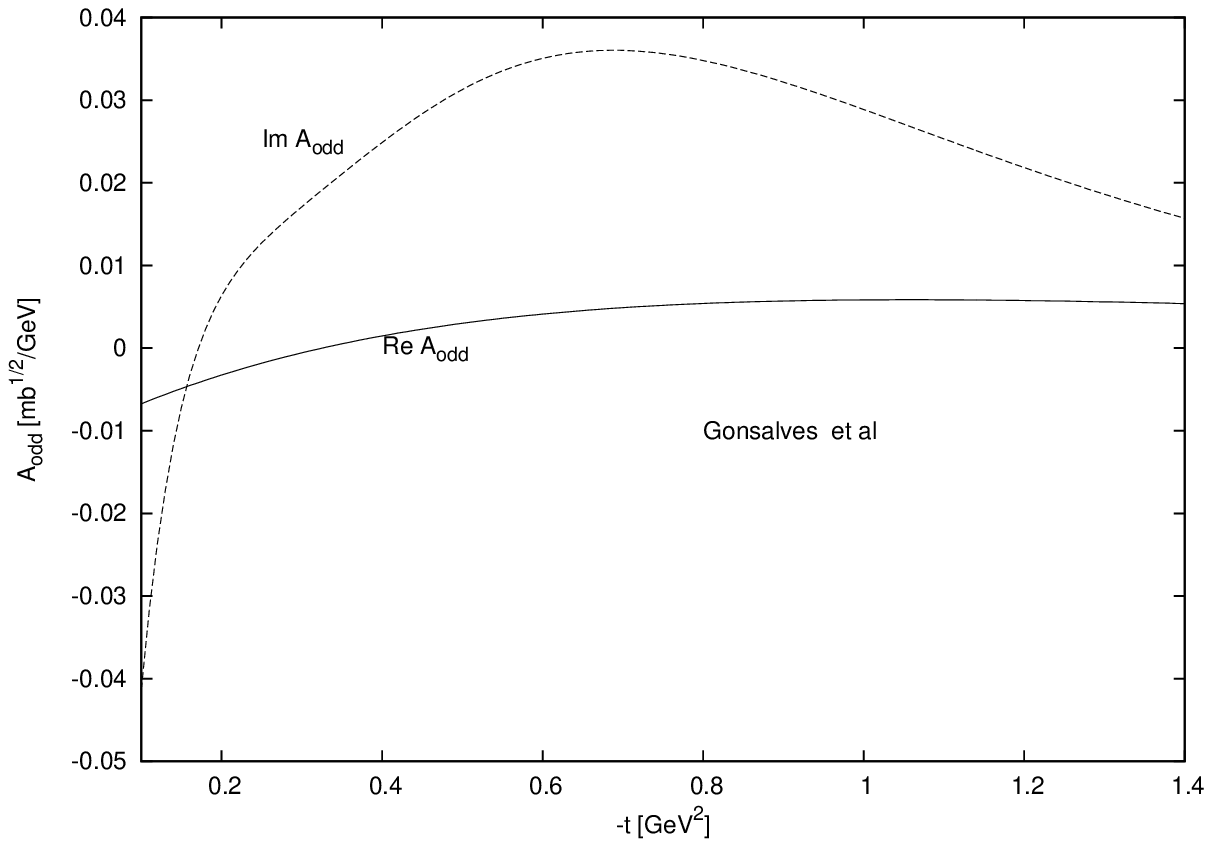, width=0.45\columnwidth }
\end{center}
\caption{The odderon amplitude extracted from the phenomenological models ~\cite{broni} (left panel) and
~\cite{silva} right panel0}
\label{fig3}
\end{figure}

Remarkably in both models
the odderon amplitude is complex (in ~\cite{broni} because the odderon is not taken as a pole in the complex $j$-plane but
rather as a dipole). The odderon amplitudes in both models are presented in Fig \ref{fig3}. Apart from a large imaginary part
both amplitudes are much greater than our calculated (real) ${\cal A}_O$ at these energies.

\begin{figure}
\begin{center}
\epsfig{file=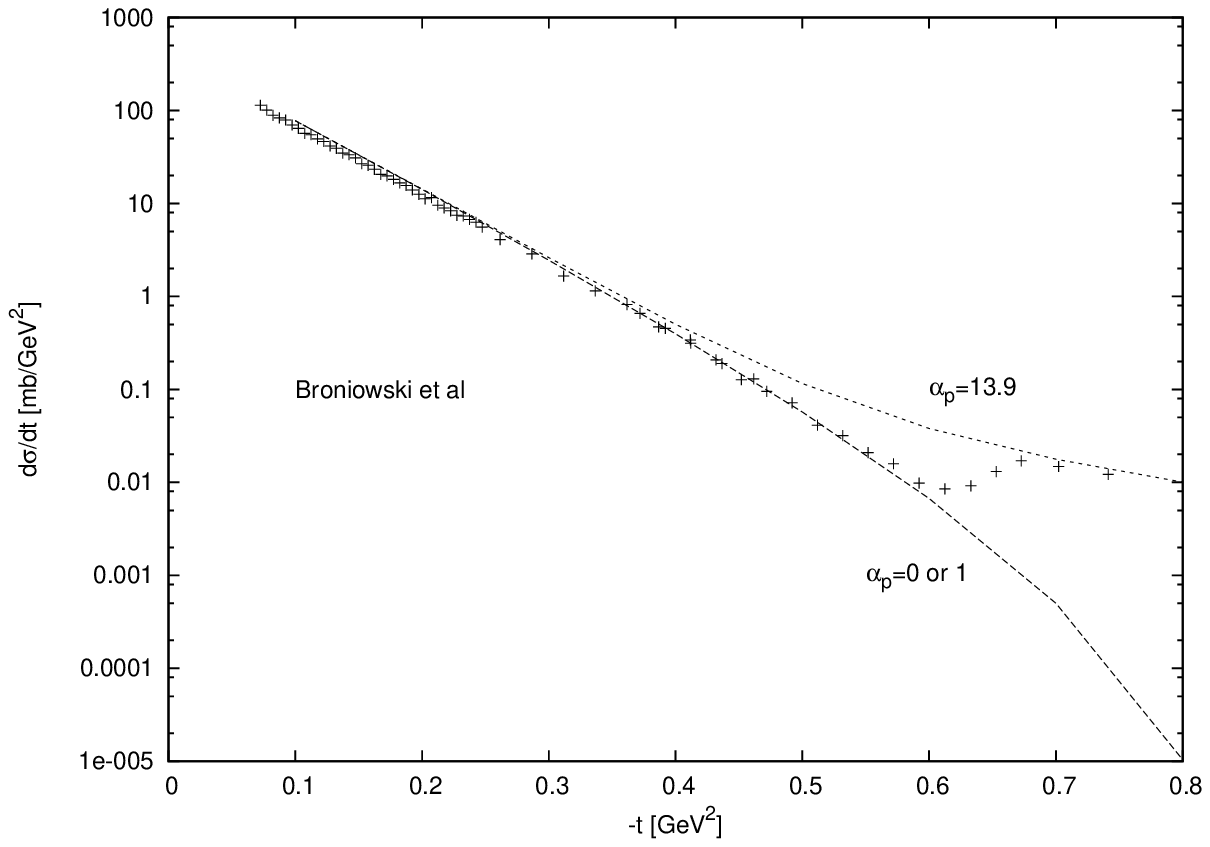, width=0.45\columnwidth }
\epsfig{file=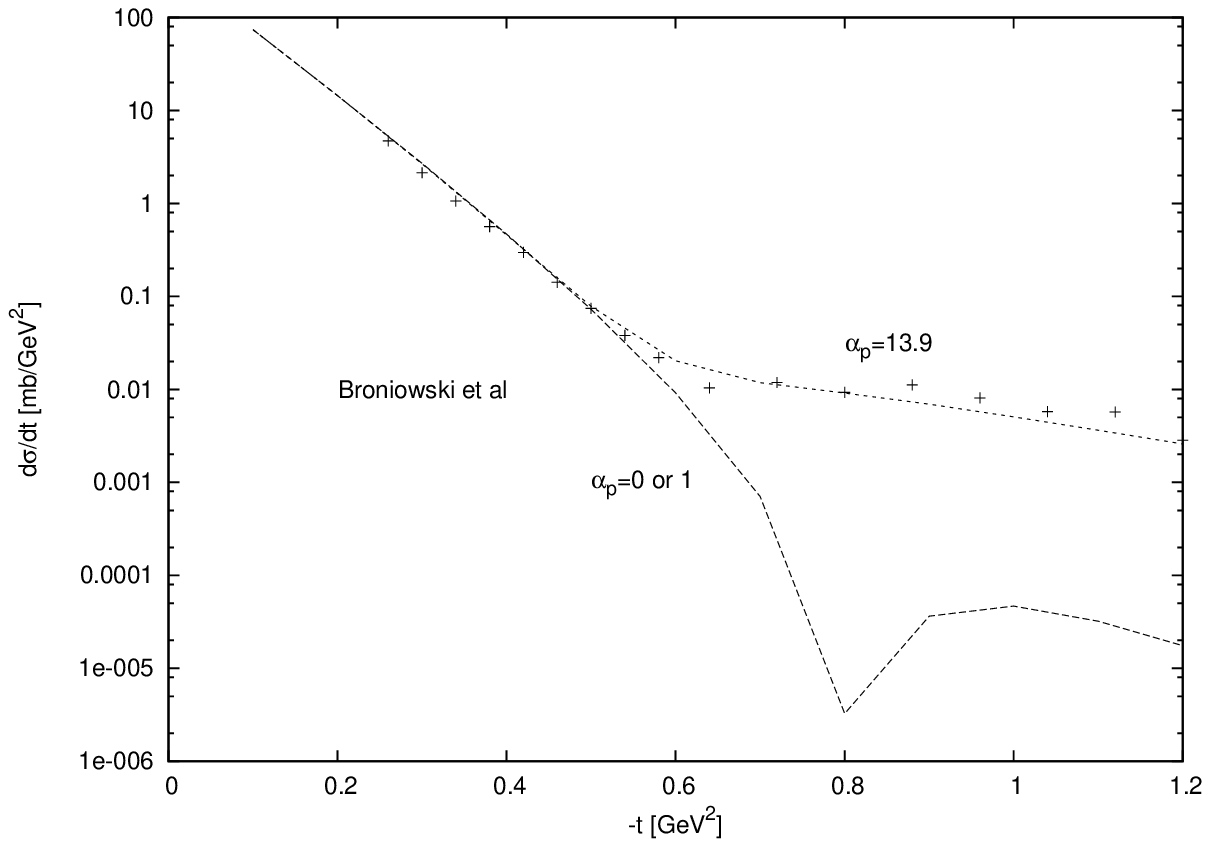, width=0.45\columnwidth }
\end{center}
\caption{The differential cross-section for elastic $pp$ (left panel) and $p\bar{p}$ (right panel)
scattering obtained after the
substitution of the $C=-1$ amplitude in ~\cite{broni} by the calculated odderon amplitude
${\cal A}_O$ at $\sqrt{s}=$2.76 and 1.96 TeV respectively. The experimental data are from ~\cite{pp276} and \cite{pbp196} respectively}
\label{fig4}
\end{figure}

\begin{figure}
\begin{center}
\epsfig{file=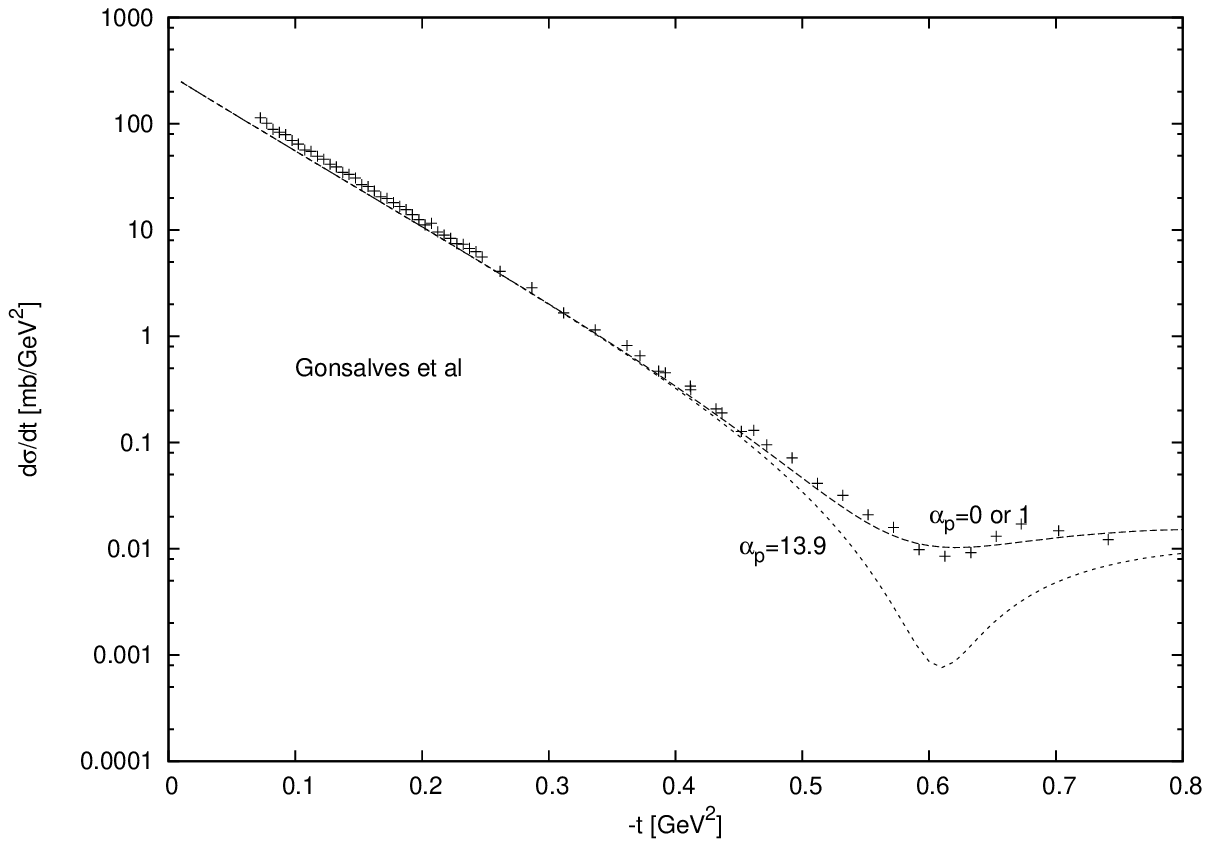, width=0.45\columnwidth }
\epsfig{file=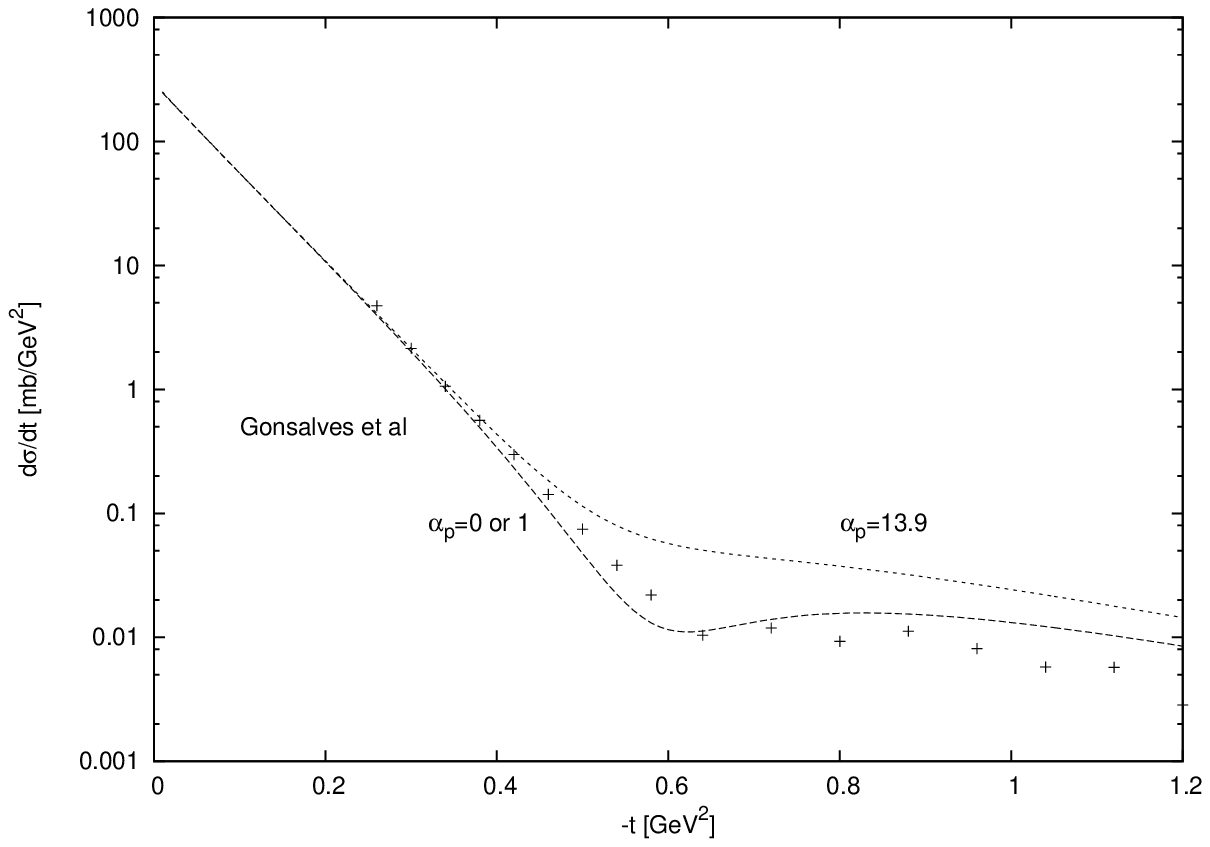, width=0.45\columnwidth }
\end{center}
\caption{The differential cross-section for elastic $pp$ (left panel) and $p\bar{p}$ (right panel)
scattering obtained after the
substitution of the $C=-1$ amplitude in ~\cite{silva} by the calculated odderon amplitude
${\cal A}_O$ at $\sqrt{s}=$2.76 and 1.96 TeV respectively.The experimental data are from ~\cite{pp276} and \cite{pbp196} respectively}
\label{fig5}
\end{figure}

So should one try to adjust to the data the cross-sections obtained after our calculated  ${\cal A}_O$  takes the role
of the odd amplitude, apart from the absence of the imaginary part, one is compelled to seriously increase its
magnitude by taking $\alpha_p$ considerably higher than unity. In the two following pictures we show our
attempts in this direction for the two models ~\cite{broni} and ~\cite{silva}. In both cases the original
amplitude with $\alpha_p=1$ (Fig. \ref{fig1}) does not practically change the cross-section without the $C=-1$ component at all,
that is with $\alpha_p=0$. In fact the difference between the curves with $\alpha_p=0$ and $\alpha_p=1$ is indistinguishable
on the adopted scale.
The results more or less in the range of the data for the parametrization of ~\cite{broni} require $\alpha_p$ in the region of $\sim 14$.
The optimal value to simultaneously describe $pp$ and $p\bar{p}$ data is 13.9 (shown in Fig. \ref{fig4}).
For the parametrization of ~\cite{silva} the situation is worse (Fig. \ref{fig5}): with the same large $\alpha_p$ one gets  a nice agreement for $pp$ but any variation of $\alpha_p$
crudely fails for $p\bar{p}$, since with the growth of $\alpha_p$ the curve moves upwards as compared to the data.

Since values of $\alpha_p$ of the order 14 do not seem physically reasonable the true result of this comparison is that the BLV odderon is simply too small
to be felt in $pp$ and $p\bar{p}$ elastic scattering. Apart from this large (and different) imaginary parts in the phenomenological $C=-1$
amplitude do not seem to be cleartly understandable from the theoretical point of view and require  explanation.

\section{Conclusions}
We have calculated the $C=-1$ amplitude corresponding to the exchange of the BLV odderon with the maximal intercept equal to exactly unity.
This amplitude is real and shows a rather peculiar $t$ dependance (Fig (\ref{fig1}). At $t=0$ it is equal to zero.
Compared to the amplitude coming from the interchange of three non-interacting gluons in the $C=-1$ state our calculated amplitude is
$\sim 1000$ times smaller with the same coupling constant. In the existing phenomenological models the $C=-1$ amplitude is complex and
about 200 times larger in magnitude. So if one believes in these models the BLV odderon with a reasonable values for the coupling constant
is far smaller to manifest itself in the $pp$ and $p\bar{p}$ elastic scattering. At $t=0$ the BLV odderon doles not contribute to the ratio $\rho$.

In fact this conclusion is not unexpected. In the processes like $\gamma^*+p \to \eta_c+p$ the cross-sections come exclusively from
the BLV odderon exchange. Previous calculations found  that these cross-sections were  extremely small,
far beyond our present experimental facilities.

The remaining open question is the role of the JW odderon with all three reggeons at different spatial points. It does not contribute
to $\gamma^*+p \to \eta_c+p$ but certainly does in the elastic $pp$ and $p\bar{p}$ scattering. It is possible that its contribution is much
greater than of the BLV
odderon. So although theoretically it diminishes with energy much stronger that the BLV odderon, the dominance of the latter is not effective at presently achieved energies
and the JW odderon can be discovered in (anti)proton elastic scattering.. However the JW odderon is an object much more complicated than the BLV odderon.
Its  Green function is not known at present. So, although very important,  its study is postponed for future investigations.

\end{document}